\documentclass[aip,jcp,
amsmath,amssymb,
preprint,%
 floatfix,%
]{revtex4-1}

\usepackage{graphicx}
\usepackage{dcolumn}
\usepackage{bm}
\usepackage[mathlines]{lineno}

\usepackage[utf8]{inputenc}
\usepackage[T1]{fontenc}
\usepackage{mathptmx}
\usepackage{etoolbox}

\begin{document}


\title[Electron thermalization length in solid para-hydrogen at  low-temperature]{Electron thermalization length in solid para-hydrogen at  low-temperature}

\author{A. F. Borghesani}
\email{armandofrancesco.borghesani@unipd.it}%
\affiliation{ 
CNISM unit, Department of Physics \& Astronomy, Università degli Studi di Padova \\and Istituto Nazionale Fisica Nucleare, sez. Padova, Padua, Italy
}

\author{G. Carugno}%
\affiliation{Istituto Nazionale Fisica Nucleare, sez. Padova, Padua, Italy}%

\author{G. Messineo}\affiliation{Istituto Nazionale Fisica Nucleare, sez. Ferrara, Ferrara, Italy}
\author{J. Pazzini}
\affiliation{Department of Physics \& Astronomy, Università degli Studi di Padova and \\ Istituto Nazionale Fisica Nucleare, sez. Padova, Padua, Italy}

\date{\today}

\begin{abstract}
We report the first ever measurements of the thermalization length of low-energy electrons injected into solid para-hydrogen at a temperature \(T\approx 2.8\,\)K. 
The use of the pulsed Townsend photoinjection technique has allowed us to investigate the behavior of quasi-free electrons rather than of massive,  slow negative charges as reported in all previous literature. We have found an average  thermalization length \(\langle z_0\rangle\approx 260\,\)\AA\ which is 3 to 5 times longer than that in liquid helium at the same temperature.
\end{abstract}

\maketitle

\section{Introduction}\label{sec:intro}
An active field of research is the measure of
the permanent electric dipole moment of the
electron 
  to look for CP violation phenomena beyond those included in the standard model of particle physics~\cite{Vutha2010,Aggarwal2018,ACMEcollaboration}.
Barium monofluoride (BaF) is a promising polar molecule to be used for this goal because its outermost electron in the HOMO is acted upon by an intrinsic electric field whose strength is otherwise unattainable in laboratory~\cite{Sandars1965}. 

The 
 PHYDES collaboration aims at embedding BaF molecules in a solid matrix of para-hydrogen (p-H\(_2\)) at cryogenic temperatures or in other cryogenic noble gas solid environments.  This matrix isolation technique limits the host-guest interaction, enhances the density of the embedded molecules, and reduces their diffusion prowess~\cite{Guarise2017,Guarise2019}. 

Solid para-hydrogen at low temperature is a matrix of great interest because of its remarkable properties due to its quantum nature. The antiparallelism of the nuclear spins leads to a configuration of lowest energy. 
The solid crystallizes in hexagonal closed-packed structure ({\em hcp} ), the average nearest-neighbor distance is \(\approx 3.8\,\)\AA\  with a molecular mean-squared displacement of \(\approx 0.48\,\)\AA~\cite{Silvera1980,Driessen1987,Meyer1998,Prisk2023}. Its number density at \(T=4.2\,\)K is fairly high, \(N\approx 2.6\times10^{28}\,\)m\(^{-3}\), and the solid can be considered to consist of an assembly of molecules all
translationally localized at lattice sites but freely rotating even in the zero-temperature limit. 

Embedding neutral BaF molecules in the solid p-H\(_2\) matrix is yet an unsolved technical problem. Actually, solid BaF\(_2\) is  evaporated in a glow discharge chamber. The BaF\(_2\) molecules are dissociated and ionized. BaF\(^+\) cations are then extracted by means of electrostatic lenses and mass-selected by a Wien filter.  Eventually, the cations are electrostatically slowed down to a final kinetic energy of \(\approx  5\,\)eV and are admitted into the low-temperature chamber where they mix with the p-H\(_2\) gas that has to be condensed to form the solid. Details on this technique will appear in a future publication. 

We plan to investigate the possibility of neutralizing the cations embedded in the solid by injecting into it electrons from a photocathode irradiated with a short UV laser pulse. To achieve this goal the knowledge of the energetics and dynamics of low-energy electrons in the solid p-H\(_2\) matrix is of paramount importance. However, to the best of our knowledge, there are only a couple of papers about excess negative charge carriers in solid p-H\(_2\)~\cite{Levchenko1990b,Levchenko1990} and a few more in solid~\cite{LeComber1976,Levchenko1988} and liquid H\(_2\)~\cite{Levchenko1992,Levchenko1994,Lerner1994,Berezhnov2003}.

Almost all the experimental papers found in literature are dealing with negative charges of unspecified nature. In those experiments ionization is obtained by either using an energetic electron beam of \(\approx 45\,\)keV~\cite{LeComber1976} or \(\beta\)-rays from Tritium sources whose average energy is in excess of \( 5\,\)keV~\cite{Levchenko1988,Levchenko1990,Levchenko1990b,Levchenko1992,Levchenko1994} whereas the dissociation energy of H\(_2\) is \(\approx 4.5\,\)eV. These energy values are sufficient for both ionization and repeated dissociation events of the  hydrogen molecules that may lead to the production of stable H\(^-\) ions or even (H\(_2\))\(_n^-\) cluster ions.

The analysis of the phenomenology of the drift mobility measurements of the negative excess charges in both liquid and solid hydrogen and in solid p-H\(_2\) convincingly demonstrates that different ionic species were detected rather than quasi-free electrons. 

In liquid H\(_2\) in the range \(12\,\mbox{K}<T<34\,\mbox{K}\,\), the value of the drift mobility \(\mu\) of negative species is comprised in the range \(1.5\times 10^{-2}\,\mbox{cm\(^2\)/Vs}<\mu<3\times 10^{-2}\,\mbox{cm\(^2\)/Vs}\), comparable to the value in liquid helium~\cite{Borghesani2007}, and shows evidence of the presence of two different negatively charged species~\cite{Sakai1983,Levchenko1992}. 
This behavior is rationalized by assuming that, owing to the very large and positive ground-state energy of an electron in the conduction band of the liquid \(V_0\approx 1.9\,\)eV~\cite{Lerner1994} and because of the strong, repulsive, short-range exchange forces both electron bubbles (i.e., electrons self-localized in empty cavities) and ionic bubbles~\cite{Volykhin1995} (i.e., H\(^-\) ions or
(H\(_2\))\(_n^-\) cluster ions surrounded by a void shell)
both coexist in the liquid~\cite{
Berezhnov2003}.

In solid hydrogen the mobility of the negative charges is very low, roughly ranging from \(4\times 10^{-7}\,\)cm\(^2\)/Vs to \(8\times 10^{-6}\,\)cm\(^2\)/Vs in the range \(12\,\mbox{K}<T<14\,\mbox{K}\)~\cite{LeComber1976,Levchenko1988}, also depending on the applied drift field, and strongly decreases with decreasing temperature. In order to rationalize these results several hypotheses have been suggested. The first one is that electrons deform the not-so-rigid solid (its surface tension is estimated to be \(\sigma\approx 4.9\times 10^{-3}\,\)J/m\(^2\) at \(T=2.8\,\)K~\cite{Corruccini1965}) and self-localize in empty bubbles giving origin to electron bubbles as was suggested for the case of solid helium~\cite{Cohen1969}, in which the mobility values are of comparable magnitude~\cite{Keshishev1971,Dahm1972,Dionne1972}. The second one is that the charge carriers are localized by the formation of small polarons and their transport occurs via phonon-assisted- or hopping  processes~\cite{LeComber1976,Shikin1977}. 
Whatever the explanation might be, it is clear from the experiments that the charge carriers cannot be quasi-free electrons.

Finally, in solid p-H\(_2\) the drift mobility of excess negative charge carriers turns out to be extremely low. 
In the temperature range \(13.6\,\mbox{K}>T>5.8\,\mbox{K}\) \(\mu\) decreases from a value \(\mu \approx 6\times 10^{-6}\,\)cm\(^2\)/Vs down to \(\mu\approx 2\times 10^{-11}\,\)cm\(^2\)/Vs~\cite{Levchenko1990b,Levchenko1990}.
Above \(T=10\,\)K the diffusion coefficient \(D\) of negative charges, obtained from the mobility by exploiting the Nernst-Debye-Einstein relationship \(D/\mu=k_\mathrm{B}T/e,\) in which \(k_\mathrm{B}\) is the Boltzmann constant and \(e\) is the electron charge, closely resembles the diffusion coefficient of ortho-hydrogen molecules and is therefore rationalized by assuming a mechanism of thermally activated diffusion of charged point defects interacting with vacancies~\cite{Andreev1989}. Below \(T=8\,\)K also a mechanism of quantum tunneling of vacancies has to be introduced. In this way the activation energy and the free energy for vacancy formation are determined and are  in good agreement with the theoretical estimate~\cite{Ebner1972} and with their determination in NMR experiments~\cite{Zhou1990}. Moreover, the authors warn that some differences in the experimental outcome are to be attributed to the anisotropic crystal structure~\cite{Levchenko1990b}.

The nature of the negative charge carriers has been identified with localized point charge defects that interact with the vacancies in the crystal~\cite{Levchenko1990b}. It is very plausible that such entities are massive H\(^-\) anions or the even more massive (H\(_2\))\(_n^-\) cluster ions as possible products of the technique used to inject charges in the solid.
Moreover, the presence of H\(_2^-\) anions has been observed in \(\gamma\)-ray irradiated, pure solid p-H\(_2\) in high-resolution ESR experiments~\cite{Kumada1996}, whereas electron bubbles are produced in place of H\(_2^-\) anions 
 only if the solid p-H\(_2\) contains
a small amount of deuterium~\cite{miyazaki2002}.
Actually, similar mechanisms of interaction between charge carriers and vacancies have been observed in solid deuterium, in which negative charges are assumed to be electron bubbles~\cite{Trusov1996}.

The whole wealth of experimental results and their interpretation strongly indicates that until now the behaviour of quasi-free electrons in fluid and solid hydrogen, including solid p-H\(_2\), has not been investigated yet. Therefore, we have carried out an experiment in which bunches of low-energy electrons are photoinjected into the solid by means of short pulses of a UV laser. We report in this paper the results of the measurements of the thermalization length of such quasi-free electrons in solid p-H\(_2\) at \(T\approx 2.8\,\)K.

The paper is organized as follows. In Sect.~\ref{sect:expdet} details on the experimental setup and procedure are given. In sect.~\ref{sect:expres} we present and discuss the experimental results. Finally, some conclusions are drawn in Sect.~\ref{sect:conc}.

\section{Experimental details\label{sect:expdet}}

A thorough description of the whole apparatus will appear in a subsequent paper. Here we only give a brief and schematic description of the experimental setup and technique.

	The measuring cell is located in a cryostat and is cooled down to  \(T\approx 2.8\,\)K using a commercially available cryocooler. The cell can be optically accessed through  suitable viewports in the cryostat and is evacuated down to a residual pressure 
	\(P_\mathrm{res}\approx 10^{-6}\,\)Pa. 	
	
	Normal hydrogen gas is enriched in the para-hydrogen fraction by exploiting standard techniques~\cite{Tom2009,Sundararajan2016,Bhandari_2021}. The gas is precooled to \(T\approx 77\,\)K by slowly flowing it in a capillary immersed in a liquid N\(_2\) bath. 
	Subsequently, the gas is flown through a copper tube, kept at \(T\approx 22\,\)K by means of an ancillary cryocooler. The tube is packed with   Fe\(_2\)O\(_3\) fine powder that acts as a catalyst for the ortho\(\rightarrow\)para conversion. The gas flow rate is controlled by a leakage valve and measured by a flow meter.
	The  gas enriched in the para fraction is then admitted into the measuring cell at low pressure. During the growth of the solid p-H\(_2\) film  the gas pressure is about \(P\approx 7\times 10^{-5}\,\)Pa. 
	
\begin{figure}[ht!]
	\centering
	\includegraphics[width=\columnwidth]{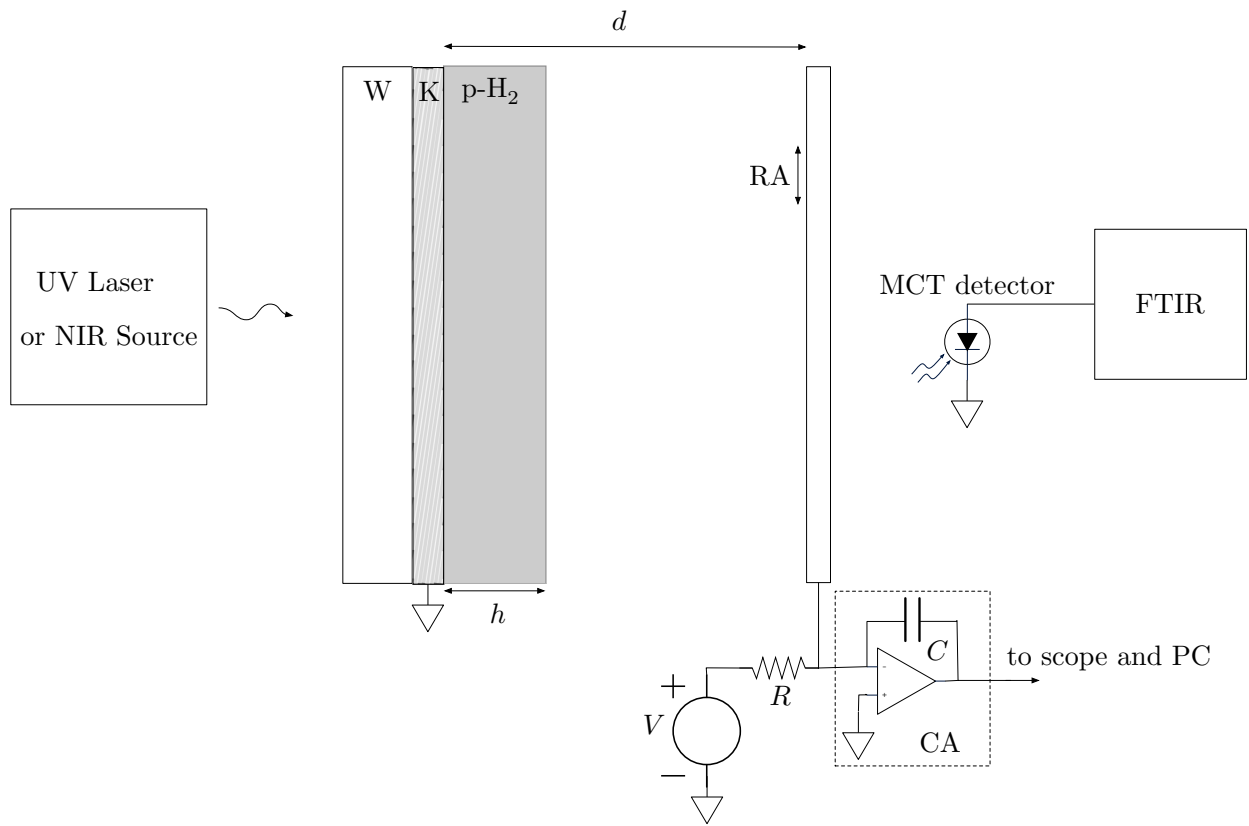}\caption{\small Conceptual scheme of the setup for measuring both the thickness of the p-H\(_2\) solid film during its growth  and the amount of photoinjected charges after the film growth is stopped. Symbol meaning: W = optical substrate. K = \(100\,\)\AA\ thin gold cathode, p-H\(_2\) = para-hydrogen film of thickness \(h\). RA = retractable anode. \(d=\) drift distance. 
	 V = d.c. power supply. C = integrating capacitor. \(R=\) high-impedance resistor.  CA = charge amplifier. FTIR = Fourier-Transform InfraRed interferometer.  \label{fig:schema}}
\end{figure}

A set of windows and electrodes is located in the cell to allow us to monitor the growth and purity of the 
p-H\(_2\) film and to inject electrons in the solid film and collect them.
A conceptual, simplified, schematics of the electrodes and windows assembly is shown in Figure~\ref{fig:schema}.
 A sapphire window, coated with a \(\approx 100\,\)\AA\ thick Au film, is mounted in the cell so as to achieve good thermal contact with it. The thin Au film is grounded and acts as the photocathode in a diode configuration of the electrodes.
 
At regular time intervals the thickness of the growing p-H\(_2\) film is monitored by recording its near-infrared (NIR) absorption spectrum, as described in literature~\cite{Fajardo2011,Fajardo2019}. The radiation of the near-infrared source (Thorlabs, model SLS202L/M) impinges on the back of the window, passes through the p-H\(_2\) film and is detected by a MCT detector (Hamamatsu, mod. P7163) whose output is amplified and fed to the FTIR interferometer (Bruker, mod. Equinox 55), thereby yielding the absorption spectrum \(I(\tilde \nu)\) as a function of the wavenumber \(\tilde \nu\) in cm\(^{-1}\).

\begin{figure}[ht!]
	\centering
\includegraphics[width=\columnwidth]{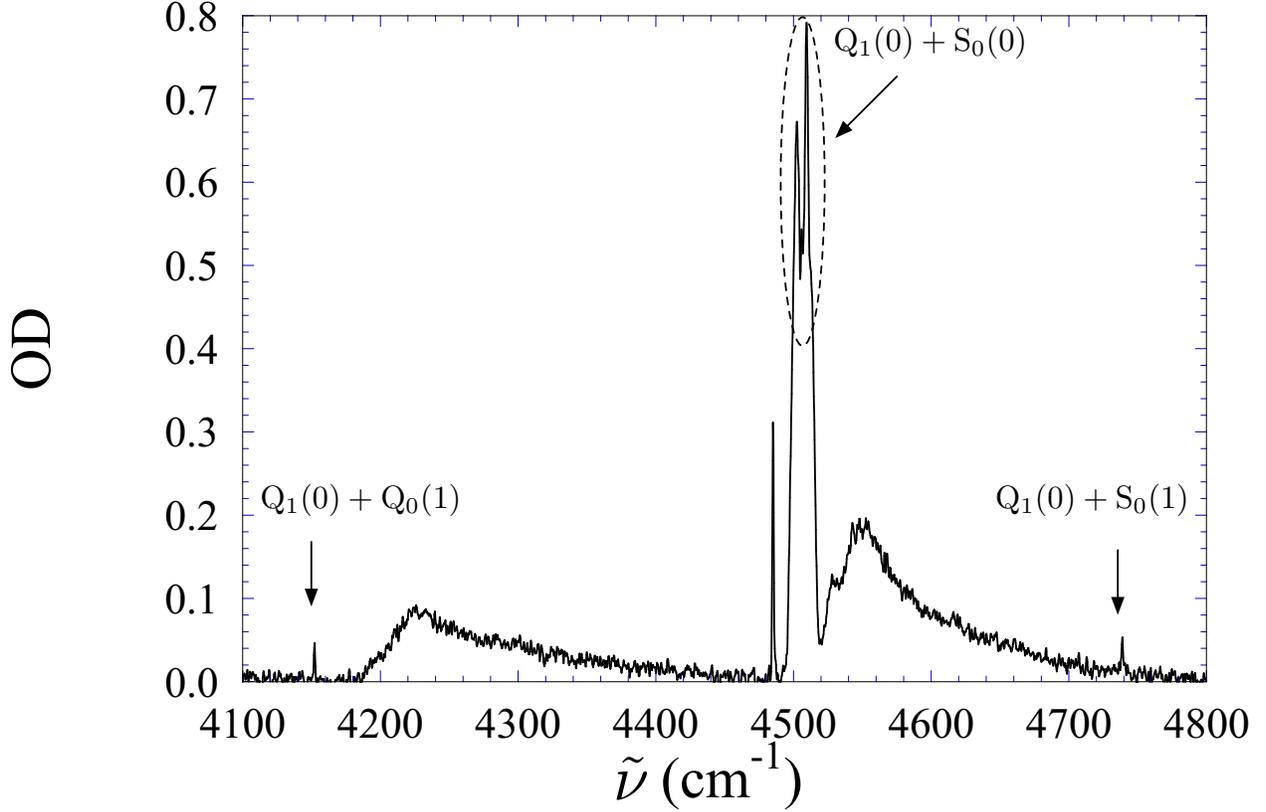}
\caption{\small Optical density spectrum of the solid p-H\(_2\) film after \( 3\,\) hrs condensation. The ro-vibrational transitions used for computing the film thickness are enclosed in the dashed ellipse. The lateral bands are analyzed to obtain the ortho-H\(_2\) concentration. The meaning of the notation of the ro-vibrational transitions can be found in literature~\cite{Bhandari_2021}.
\label{fig:OD}}
	\end{figure}

 According to literature the film thickness \(h\) and the ortho-H\(_2\) fraction \(f\) can be computed from the optical density \(\mbox{OD}= -\log\left[I(\tilde\nu)/I_0(\tilde \nu)\right]\), in which \(I_0\) is the spectrum recorded in absence of p-H\(_2\), respectively as~\cite{Fajardo2011,Fajardo2019}

\begin{equation}
h(\mbox{mm})= 0.048\times\int\limits_{4495}^{4525}\mbox{OD}(\tilde \nu)\,\mathrm{d}\tilde \nu,
\label{eq:h}
\end{equation}

\begin{equation}
	f=\frac{0.124}{h}\times\int\limits_{4151}^{4154}\mbox{OD}(\tilde \nu)\,\mathrm{d}\tilde \nu,\quad \mbox{or } f=\frac{0.0787}{h}\times\int\limits_{4732}^{4742}\mbox{OD}(\tilde \nu)\,\mathrm{d}\tilde \nu
\label{eq:orthoF}
\end{equation}
A typical OD spectrum recorded after \(\approx 3 \,\)hrs condensation is reported in Figure~\ref{fig:OD}.
In all experimental runs we obtained a constant growth rate of \(\approx 100\,\mu\)m/hr and a 
 constant ortho fraction \(2\,\%<f<3\,\%\).  

\noindent After the solid film thickness has reached a value between \(500\) and \(600\,\mu\)m the gas inlet is closed and the film growth is almost stopped, its growth rate dropping down to \(\approx 0.5\,\mu\)m/hr, whereas the ortho-hydrogen content remains constant for a longer time interval after the photoinjection experiment has been completed.

Once the solid film has reached the desired thickness, the retractable anode is set in position in front of the cathode by means of a suitable manipulator.  The NIR source is replaced by a 266-nm pulsed laser. The laser, whose spot has a diameter \(\phi \approx 8\,\)mm, impinges on the back of the Au cathode and photoextracts electrons. The laser pulse has a duration  \(\Delta t\approx 10\,\)ns and an energy per pulse that can be varied in the range  \(50\,\mu\mbox{J}<W<150\,\mu\,\)J, but is kept constant during the measurement. The laser pulse repetition rate is set at \(10\,\)Hz.

A d.c. power supply energizes the drift capacitor up to \(1\,\)kV and sets the desired value of the electric field in the drift region of length \(d=1\,\)cm.
The current induced in the circuit by the drifting photoextracted electrons is integrated by a charge amplifier (i.e., an active integrator) whose time constant is \(\tau\approx 400\,\mu\)s and whose output amounts to \(\approx 0.25\,\)mV/fC.
The charge amplifier output is connected to a digital scope. For each setting of the drift field, 2000 signals are averaged together in order to improve the signal-to-noise ratio. The averaged signals are then fetched by a PC for offline analysis.

\section{Experimental results and discussion\label{sect:expres}}
In this Section we report the experimental results and their rationalization.

\subsection{Characterization of photoextraction in vacuo\label{sect:vacuo}}
\begin{figure}[ht!]
\centering\includegraphics[width=\columnwidth]{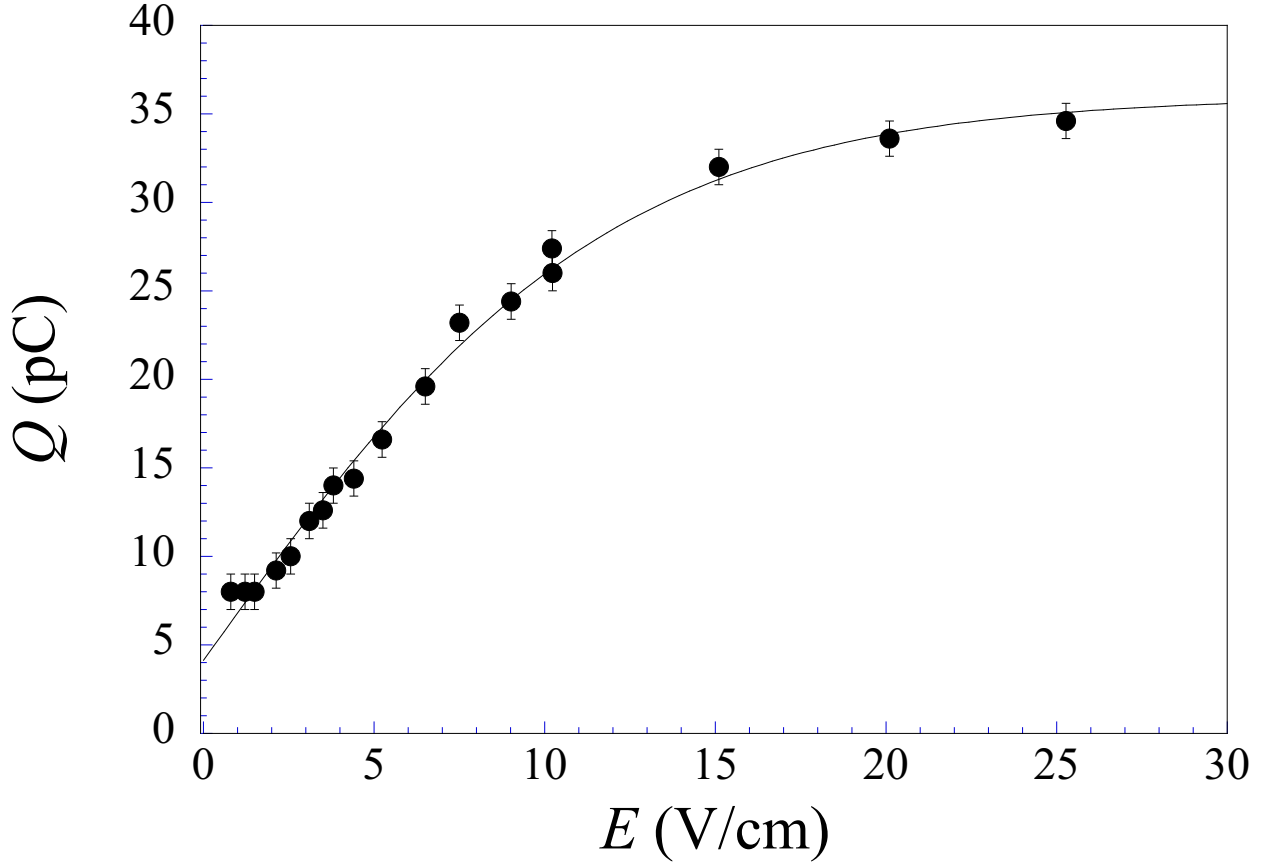}
\caption{\small Electric field dependence of the charge collected at the anode {\em in vacuo}. The solid is only a guide for the eye.\label{fig:vuoto}}
\end{figure}

First of all, the behavior of the system has to be characterized {\em in vacuo}.
We report in Figure~\ref{fig:vuoto} the charge photoextracted under vacuum conditions.
The amount of photoextracted charge increases with increasing field, as expected, and saturates for an electric field strength as weak as \(E\approx 15\,\)V/cm, when practically the whole amount of electrons produced by the laser pulse is collected.

The energy distribution of the photoelectrons is not known, as is the work function of the deposited Au film. We directly measured the highest energy of the electrons by reversing the electric field polarity in order to nullify the collected charge, thereby obtaining \(\epsilon_\mathrm{max} =(0.56\pm 0.05)\,\)eV. 
The amount of the collected charge proportionally varies with the laser pulse energy, its electric field dependence remaining the same.

\subsection{Measurements with solid p-H\(_2\) in place\label{sect:pH2}}

The measurements of photoextraction have been carried out after the thickness \(h\) of the solid p-H\(_2\) film reached a value  between \(500\,\mu\)m and \(600\,\mu\)m.
If the applied potential difference between anode and cathode is \(V\), the electric field strength \(E\) in the dielectric film is given by\begin{equation}
E\,(\mbox{V/cm})=\frac{V}{K(d-h)+h}\approx 0.792V\label{eq:E},	
\end{equation}
in which \(V\)  is measured in Volt, \(d=1\,\)cm is the drift distance and \(K\approx 1.28\) is the relative dielectric constant of solid p-H\(_2\)~\cite{Younglove1968,Constable1975}. An accurate determination of \(h\) is not extremely important as a \(20\,\%\) uncertainty in \(h\) brings about a negligible \(0.2\,\%\) uncertainty in \(E.\)
The electric field strength in the p-H\(_2\) vapor is \(E_v\,(\mbox{V/cm})=KE\approx1.01V\,\mbox{(V)}.\) We have investigated the range \(10\,\mbox{V/cm}\lesssim E\lesssim 600\,\mbox{V/cm}\).
\subsubsection{Signal waveform analysis}
\begin{figure}[ht!]
\centering
\includegraphics[width=\columnwidth]{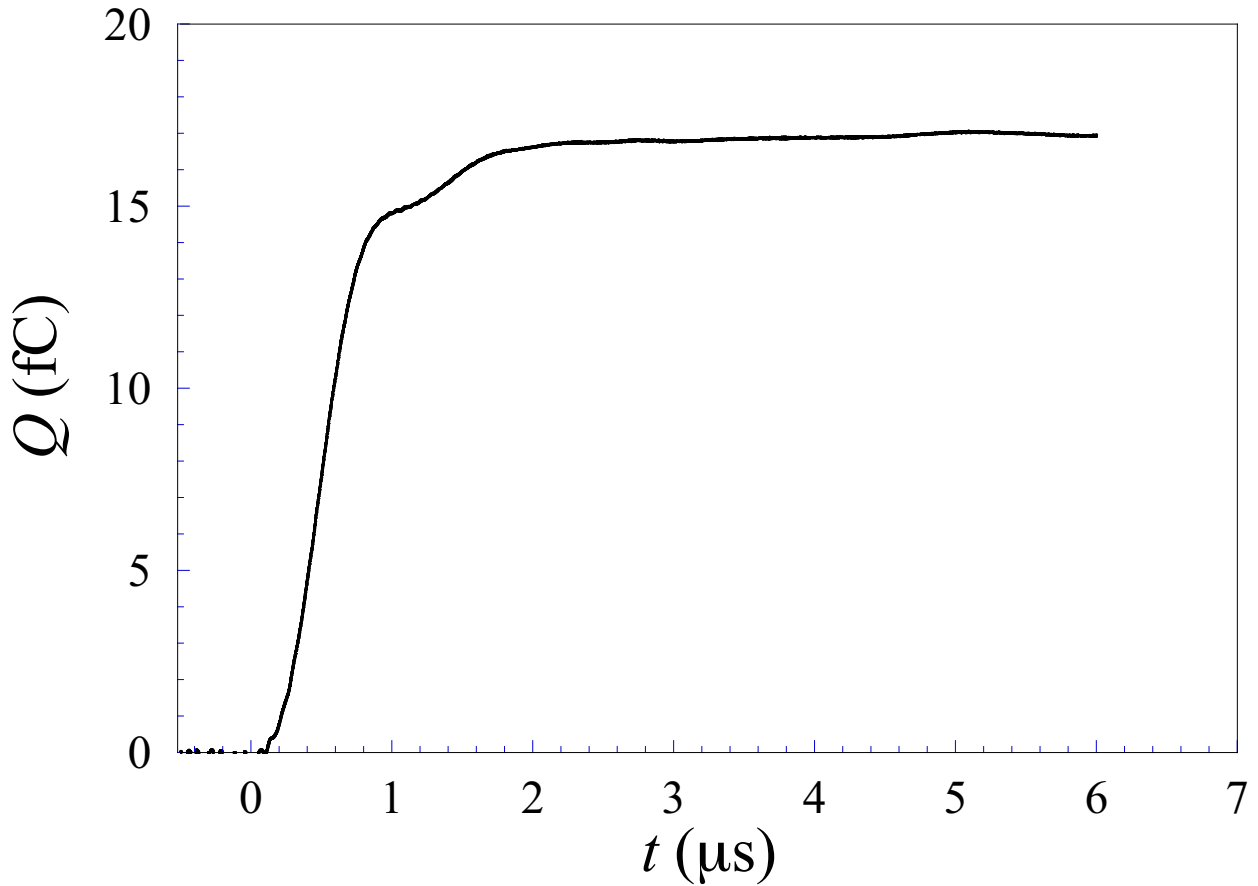}
	\caption {\small Time evolution of the collected charge in presence of the solid p-H\(_2\) film recorded with \(E=280\,\)V/cm.
	\label{fig:sigwavfor}}
\end{figure}

Some preliminary pieces of information can be gathered by inspecting the acquired signal waveforms, such as the one shown in Figure~\ref{fig:sigwavfor}. 
First of all, we note that the amount of collected charge is a factor more than 2000 smaller than that collected {\em in vacuo} with a much smaller electric field strength. We will discuss this reduction later in greater detail.   For now, we want to stress the fact that the signal rise time \(\Delta t\simeq 0.7\,\mu\mbox{s}\) is independent of the applied electric field and is solely due to the response of the electronics. If electrons injected into the solid were prone to dissociate H\(_2\) to form H\(^-\)  anions or to form electron bubbles, their drift time through the solid would be no shorter than several tens of seconds. However, we do not observe any charge pile-up in the solid in spite of the quite high laser pulse repetition rate.

 On the other hand, the transit time of electrons emerging into the vapor is estimated to be several orders of magnitude smaller than the rise time of our electronics because of the extremely low density of the vapor at \(T\approx 3\,\)K. 
 Unfortunately, there are no measurements of electron drift velocity \(v_D\) in p-H\(_2\) gas for so high an electric field strength and at so low a temperature except those  at \(T=76.8\,\)K,  at a lowest pressure \(P\sim 500\,\)Pa, and at the highest reduced field \(E/N\sim 2.6\,\)Td (\(1\,\mbox{Td} = 10^{-21}\,\mbox{V\,m}^2\)), yielding a drift velocity \(v_D\approx 3.3\times10^6\,\)cm/s~\cite{Crompton1971,Robertson1971}.
 Just to carry out a back-of-the-envelope estimate, we assume that the pressure in our experiment is so low that the vapor can be treated as an ideal gas because the equation of state of p-H\(_2\) is only known for \(T>13.8\,K\)~\cite{Leachman2009}. 
  We, thus, get a gas number density 
  \(N\approx 2\times 10^{18}\,\)m\(^{-3}\), thereby yielding 
  \(E/N\approx 5\times 10^5\,\) Td for \(E=10\,\)V/cm. As there are no hints at how fast electrons travel at such huge a reduced field strength, we just assume that they move {\em in vacuo} and obtain a maximum final 
   speed \(v\sim 2\times 10^8  \,\)cm/s, thereby leading to a transit time \(\tau\sim 5\,\)ns, much shorter than the rise time of the electronics.
  
  This analysis of the waveform leads to the conclusions that: i) electrons are actually injected into the solid in which they move at high speed, and ii) neither low mobility electron bubbles nor anions form.

  \subsubsection{Electric field dependence of the photoextracted charge\label{sect:qvse}}
\begin{figure}[htp!]
	\centering\includegraphics[width=\columnwidth]{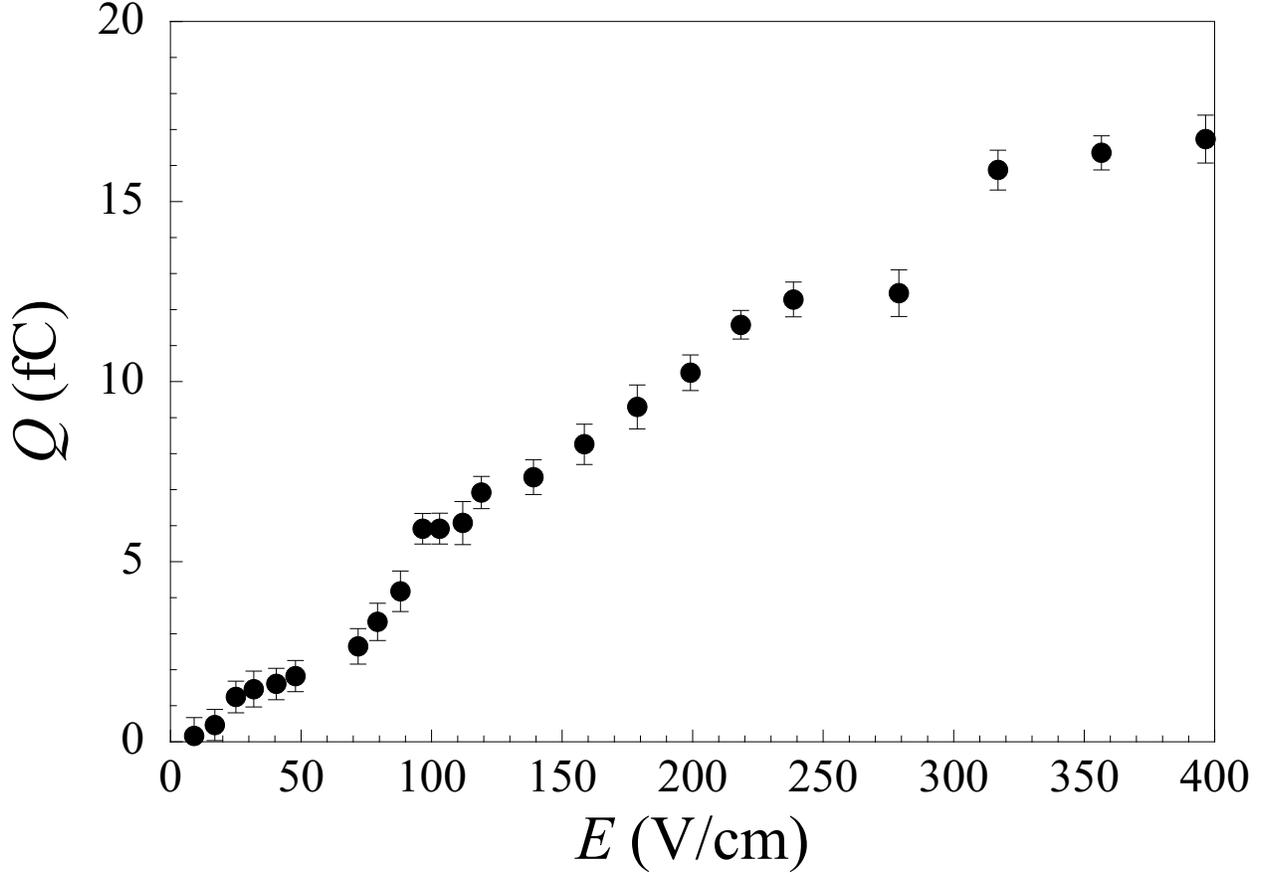}
	\caption{\small Collected charge \(Q\) vs the electric field \(E\) acting on the p-H\(_2\) film. \label{fig:QvsE}}
\end{figure}

The typical outcome of the experiment is shown in Figure~\ref{fig:QvsE}, in which the amount of the charge collected at the anode is plotted as a function of the electric field \(E\) in the p-H\(_2\) film.

As previously noted, the electrons photoinjected into the solid p-H\(_2\) film are epithermal. Upon injection electrons  will lose their excess energy by collisions  in the solid until they get thermalized and then propagate 
in the conduction band of the crystal. The model used to describe the injection process has been thoroughly described in literature~\cite{Onn1969,Onn1971,Smejtek1973}. We briefly recall here its main features.

The electrons injected in the solid film close to the Au/p-H\(_2\) interface are under the influence of their image force, which acts so as to push them back to the cathode, and of the externally applied electric field \(E\), which pulls them towards the collector. The potential energy (in eV) of the electrons  is given by
\begin{equation}
	\varphi(z)=V_0-{A}{z}^{-1}-
	Ez\quad\mbox{for }z\ge 0.
	\label{eq:pot}
\end{equation}
Here, \(z\) is a coordinate orthogonal to the Au/p-H\(_2\) interface where the coordinate origin is located, and  \(V_0\) is the electron ground state energy at the minimum of the conduction band in the solid.  The coefficient \(A\) takes into account the dielectric properties of the solid film and is given by~\cite{Griffiths}
\begin{equation}
	A=\frac{e
	}{16\pi\epsilon_0 K}\left(\frac{K-1}{K+1}\right)\approx 0.345
	\,\mbox{eV\,\AA}.
\label{eq:A}	\end{equation}
	\(\varphi(z)\) has a maximum at \(z_m=\sqrt{A/E}\) where the Schottky barrier lowering (in eV) amounts to \(\Delta \varphi=-2Ez_m\).  In our experimental conditions 
	\( 200\,\mbox{\AA\ }\lesssim z_m\lesssim 2000\,\mbox{\AA} \), whereas \(4\times 10^{-5}\,\mbox{eV}\lesssim\vert \Delta\varphi\vert \lesssim 3\times 10^{-4}\,\mbox{eV}\) is negligible.

	We note that  the distance of the potential maximum from the Au/p-H\(_2\) interface is always much shorter than the solid film thickness, \(z_m\ll h\). 
	However, we cannot be sure that \(z_m\) is  greater or comparable to the electron mean free path (mfp) \(\ell\)  in the solid. We can reasonably assume that this is the case. Actually, on one hand, if we could make the hypothesis that the molecules in the solid were a collection of independent scattering centers, we would get \(\ell\approx 4\,\mbox{\AA}\) by using the recommended value for the e-H\(_2\) scattering cross section at low energy (\(\sigma_s\approx 9.2\,\)\AA\(^2\))~\cite{Yoon2008}. This mfp value, surprisingly close to the nearest-neighbor distance in the solid, is far too short to be reasonable. Actually, at low energies where the electron wavelength is comparable or even larger than the typical intermolecular distance, we expect deep modifications of the gas-phase behavior. Coherent scattering of the electron wave function within the solid may even deeply lower the gas-phase cross section~\cite{Sanche1988}.
	On the other hand, a mfp values of a few tens of times larger than the nearest-neighbor distance might be more realistic. In any case, we assume that \(z_m\gtrsim\ell\).
	
	A first attenuation of the amount of charge injected in the solid with respect to the vacuum case is due to the quantum reflection of the electron wavefunction at the barrier \(V_0\). This phenomenon strongly reduces the amount of charge that can be collected at the anode, as experimentally observed.
	
	Once successfully injected into the solid, an electron loses its excess energy by scattering off  molecules, or phonons, or structural defects. Thus, electrons undergo a random walk through the medium with the possibility that somewhere thermalization occurs due to some inelastic process (as, for instance, by inducing rotational excitation in the H\(_2\) molecules).
	Several possible paths for electrons can be singled out.
	The first possibility is that the electron is backscattered towards the cathode at its first collision~\cite{YB1933} or may backdiffuse towards the cathode before thermalization occurs. These two processes do not depend on the position of the potential maximum. 
A second possibility is that the electron thermalizes for \(z<z_m\) and backdiffuses towards the cathode.  All these processes lead to a reduction of the amount of charge that can be collected at the anode. 
The third and final possibility is that thermalization occurs for \(z>z_m\) and the electron is collected at the anode because of the pull exerted by the external field. In some sense, the potential maximum acts on the electrons as a gate at a distance \(z_m\) from the cathode. 

By taking into account all the aforementioned effects, it has been shown that, at low temperature and high density,  the amount of charge injected into the solid and collected at the anode
is accurately described by the following equation~\cite{Onn1969,Silver1970,Onn1971}

\begin{equation}
	\frac{Q}{Q_0}=\frac{B}{D}\mathrm{e}^{-z_m/z_0},
	\label{eq:Q}
\end{equation}
in which \(Q_0\) is the amount of charge that would be collected {\em in vacuo}, and \(z_0\) is the thermalization length. \(B\) is a coefficient that describes the effect of the barrier and could be computed if the energy distribution function of the photoelectrons were known~\cite{Broomall1976,Gushchin1982}. In our case, \(B\) is constant because the energy distribution function of the photoextracted electrons does not change from run to run. \(D\) is a term that describes the backdiffusion process. It can be written as \(D=1+Cz_0/\ell\), where \(C\) is a numerical constant, and does not depend on \(z_m\).
Moreover, as all the measurements with the solid film in place have been carried out for \(E\gtrsim \,15\,\)V/cm, \(Q_0\) can also be considered as a constant.

\begin{figure}[h!]
\centering\includegraphics[width=\columnwidth]{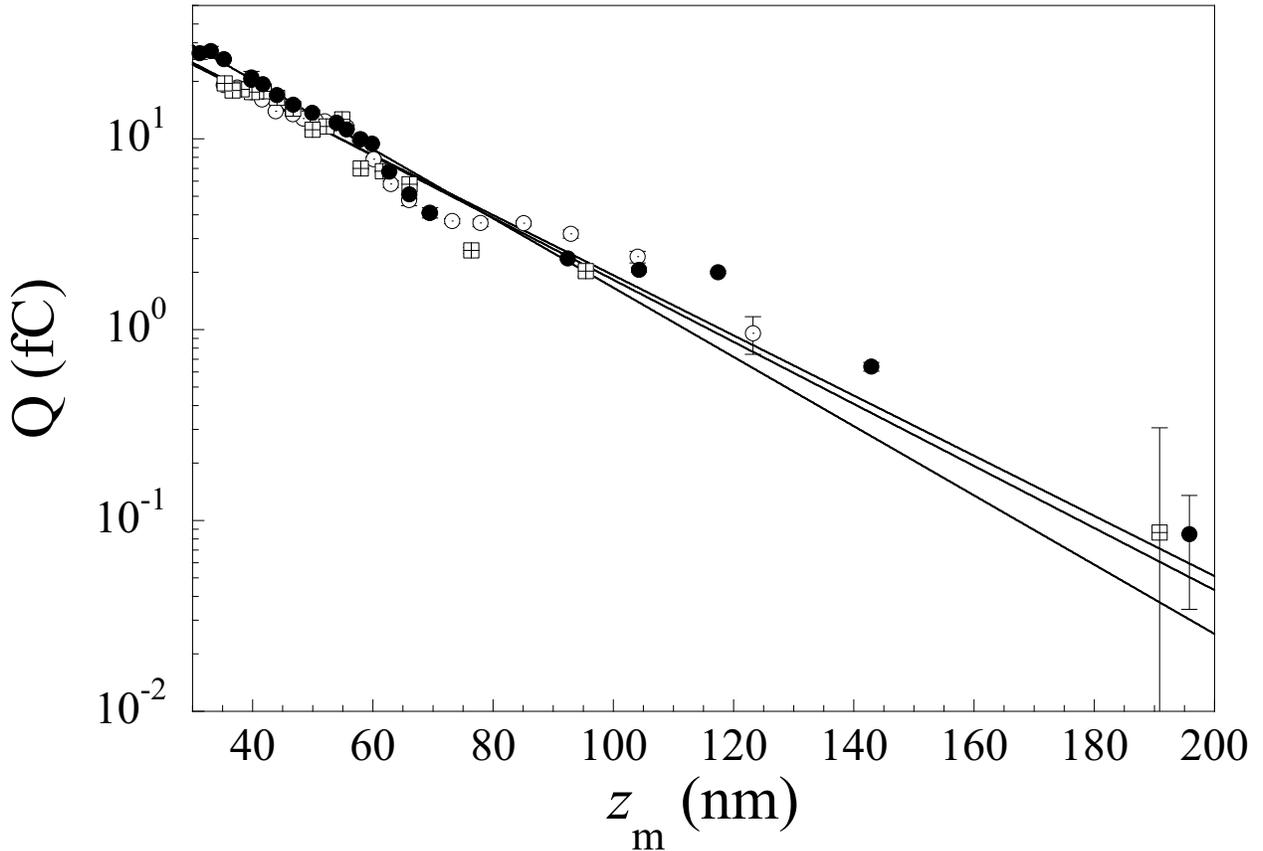}
\caption{\small  The charge collected at the anode \(Q\) as a function of the coordinate \(z_m\) of the potential energy maximum measured in the three  experimental runs. The charge in each run has been normalized to the laser energy. The solid lines are the best fit curves based on  Eq.~(\ref{eq:Q}).\label{fig:3runs}}
\end{figure}

In Figure~\ref{fig:3runs} 
we report the experimental results for the collected charge as a function of the coordinate \(z_m\) of the potential energy maximum obtained in the three different runs we have carried out at roughly one month interval from each other. The charge amount in each run has been normalized to the laser energy.
	
The experimental data satisfactorily follow 
the exponential law Eq.~\ref{eq:Q}. We obtain three independent determinations of the thermalization length that are reported in Table~\ref{tab:zmvalues}.
\begin{table}[ht!]
\begin{center}\begin{tabular}{|c|c|c|}\hline \mbox{run} & $z_0$ (\AA)  & \mbox{uncertainty} (\AA) \\\hline 1& 256 & 20 \\\hline 2 & 291 & 28 \\\hline 3 &257 & 12 \\\hline \end{tabular} \caption{Thermalization length determinations in the three experimental runs with the relative uncertainties.\label{tab:zmvalues}}
\end{center}\end{table}
The three different determinations of the thermalization length are in good agreement with each other, thereby yielding a weighted average 
\begin{equation}\langle z_0\rangle =(261\pm 10)\,\mbox{\AA}\label{eq:z0}.\end{equation}
We note that the ratio of the thermalization length to the film thickness is \(d/z_0\sim 2\times 10^4\). This high value of the ratio ensures that boundary effects are negligible,
 thereby also lending credibility to the statistical approach leading to Eq.~\ref{eq:Q}.
 
The present value of the thermalization length \(\langle z_0\rangle \) turns out to be 3 to 5 times longer than in liquid helium at approximately the same temperature~\cite{Onn1969}. 

A lack of knowledge of the electron drift mobility and/or scattering cross section in the solid p-H\(_2\) at the low temperature of our experiment does not allow us to gather some deeper pieces of information regarding the electron thermalization time that may be useful for the design of the experiment aimed at the neutralization of cations embedded in the solid.

Nevertheless, we conclude that thermalization of quasi-free electrons occurs well within the bulk. Upon thermalization, the electron thermal speed at \(T=2.8\,\)K is \(
v_\mathrm{th} \approx 11\times 10^3\,\mbox{m/s}
\)
to be compared to the average value of the sound speed \(v_S\approx 2\times 10^3\,\)m/s~\cite{Kuroda2001,Bezuglyi1971,Wanner1972,Wanner1973}. Thus, it is reasonable to assume that polarization of the medium via phonons is too slow to respond to the electron motion, which could otherwise  hinder the Coulomb interaction of the electrons with the embedded BaF\(^+\) cations.

\section{Conclusions\label{sect:conc}}
For the first time we have carried out measurements of the thermalization length of quasi-free electrons in solid para-hydrogen at low temperature. The goal is to investigate their energetics and dynamics in order to ascertain the possibility to neutralize cations embedded in a solid matrix of this material.

The outcome of the present experiment has several facets. First of all,
electrons of low energy are injected into the conduction band of the solid by exploiting the photoelectric effect of a gold cathode irradiated with a short pulse of a UV laser whose photon energy is just close to the dissociation energy of H\(_2\). By so doing, the production of very slow H\(^-\) anions is very unlikely. An analysis of the time evolution of the waveforms we have recorded indicates that   only fast negative charges are present.

Secondly, we have been able to determine the thermalization length \(z_0\) of quasi-free electrons. It turns out that \(z_0\) is quite large though well much shorter than the solid film thickness. This is a useful piece of information regarding the possibility of neutralizing the cations that have to be embedded in the solid p-H\(_2\) matrix in order to carry out high-precision measurements of the electric dipole moment of the electron.

However, the thermalization length of electrons in a matrix is an important quantity in several more physics fields. It is well known that there might be a relationship between electron thermalization and mobility~\cite{Sano1977}.
Moreover, among others, electron thermalization plays a role in the physics of geminate ion recombination or of free ion yield which is important in the field of radiation detectors~\cite{aprile,ferradini}, just to mention some applications.

In this first experiment we have investigated the electron thermalization at one single temperature and density. We had to assume that there is a barrier to the injection of electrons into the solid without actually measuring it because its determination requires a detailed knowledge of the density dependence of the thermalization process. Moreover, it was impossible, in the present experimental configuration, to measure the electron drift time. Therefore, we are planning for the near future to extend our measurement campaign to fill in this gap.

\section*{Acknowledgments\label{sect:ack}}
We wish to thank dr. D. Sali, Prof. A. C. Vutha, Prof. T. Momose, Prof. J. D.  Weinstein, and Prof. C. Caesar for helpful suggestions.
A. F. Borghesani gratefully acknowledges useful discussions with Prof. A. G. Khrapak, Prof. L. P. Mezhov-Deglin, and Prof. J.-P. Jay-Gerin.
 We also express our thanks to Fulvio Calaon for technical support.
 G. Messineo acknowledges financial support from the European Union’s Horizon2020 research and innovation programme under the Marie Skłodowska-Curie grant no. 754496. \section*{Author declarations}
 \subsection*{Conflict of Interest}
 The authors have no conflicts to disclose.
 \subsection*{Authors contributions}
 All authors equally contributed to the experiment.
 \subsection*{Data availability}
 The data that support the findings of this study are available from
the corresponding author upon reasonable request.
%

\end{document}